\documentclass[11pt]{article}
\usepackage[margin=1in]{geometry}
\usepackage{graphicx}
\usepackage{booktabs}
\usepackage{amsmath,amssymb}
\usepackage[numbers,sort&compress]{natbib}
\usepackage{hyperref}
\usepackage{tabularx}
\usepackage{multirow}
\usepackage{siunitx}
\usepackage{flafter}
\usepackage{placeins}
\usepackage{float}
\AtBeginDocument{%
  }

\begin{document}

\title{A Generative Framework for the Creation of Multi-Attribute Geographically-Explicit Synthetic Population}


\author{%
  Jinlin Wu\thanks{The Hong Kong University of Science and Technology (Guangzhou), Guangzhou, China. Emails: \href{mailto:jwu923@connect.hkust-gz.edu.cn}{jwu923@connect.hkust-gz.edu.cn}, \href{mailto:siqiao@hkust-gz.edu.cn}{siqiao@hkust-gz.edu.cn}, \href{mailto:yliu017@connect.hkust-gz.edu.cn}{yliu017@connect.hkust-gz.edu.cn}, \href{mailto:najiang@hkust-gz.edu.cn}{najiang@hkust-gz.edu.cn}.}%
  \quad Si Qiao\footnotemark[1]
  \quad Yi Liu\footnotemark[1]\\[0.5em]
  Fuzhen Yin\thanks{University of Colorado Colorado Springs, Colorado Springs, CO, USA. ORCID: \href{https://orcid.org/0009-0003-4981-7345}{0009-0003-4981-7345}. Email: \href{mailto:fyin@uccs.edu}{fyin@uccs.edu}.}%
  \quad Na Jiang\footnotemark[1]%
}






\maketitle
\begin{abstract}
Generating multi-attribute synthetic populations with realistic joint distributions and geographic variation is a foundational requirement for geo-simulation techniques, such as micro-simulation and agent-based modeling. However, it remains challenging for existing methods to reconstruct region-specific joint distributions from aggregated-level data alone. Thus, we propose a hierarchical diffusion-based generative framework that utilizes a realistic region-specific joint distribution of multiple attributes as the training target to create a synthetic population along with assigning their explicit home and work locations. Applied to 50 U.S. states and Washington, D.C., this framework generates a nationwide geographically-explicit synthetic population consisting of 332,387,543 individuals with five attributes (e.g., age, gender, employment, education, income). Held-out regional experiments show improved reconstruction of joint distributions relative to Iterative Proportional Fitting (IPF) and a one-shot diffusion baseline. At the same time, the location assignment preserves major residential and workplace patterns. As such, the proposed framework provides a scalable generative approach for creating geographically explicit synthetic populations at both regional and national levels. By reconstructing region-specific joint distributions of these five attributes using this framework, the resulting synthetic population could introduce more realistic behaviors into geo-simulations, such as agent-based modeling, enabling further exploration of the emergence of complex urban phenomena through human interactions.
\end{abstract}



\section{Introduction}\label{sec:intro}

Creating multi-attribute synthetic populations has long been considered foundational to geo-simulation techniques, such as micro-simulation and agent-based modeling \cite{mueller2011, barthelemy2013, la2025, chapuis2022}. To build such synthetic populations, research efforts have focused on reconstructing multi-attribute joint distributions from aggregate-level data (e.g., census tract-level data). However, because such data only record the marginal distributions of the population's characteristics in a certain region, they cannot fully reflect how population characteristics actually co-occur at the individual level. For example, when synthesizing a population with five attributes and each containing four categories, the full joint distribution comprises over one thousand potential combinations; yet, information related to how these attributes co-occur is missing from aggregate-level data. As a result, reconstructing the true joint distribution of a synthetic population's multi-attributes remains challenging. This could introduce internal demographic biases into the synthetic population, such as assigning a 16-year-old teenager as the child of a married couple aged 22. Furthermore, such internal demographic biases within the synthetic population could undermine the behavioral realism for the utilization of micro-simulation or agent-based modeling.

To address such a challenge, existing population-synthesis methods attempt to reconstruct the joint distribution of multi-attributes through two main approaches: synthetic reconstruction and combinatorial optimization \cite{chapuis2022}. In synthetic-reconstruction, methods such as Hierarchical Sampling (HS) \cite{jiang_large-scale_2024}, Iterative Proportional Fitting (IPF) \cite{predhumeau_synthetic_2023}, and Iterative Proportional Updating (IPU) \cite{ipu2009} create the multi-attributes synthetic population from aggregated level data (e.g., census tract-level data) by matching target marginals using heuristic to reconstruct the joint distribution of the synthetic population's attributes (e.g., age, gender, and household type) \cite{choupani2016,lovelace2015}. Even though, recent optimization method (e.g., integer least squares programming \cite{lin2023}) and parametric method (e.g., Markov Chain Monte Carlo \cite{farooq2013} and Bayesian-network sampling \cite{zhou_creating_2022}) are utilized to reconstruct the joint distribution of the synthetic population's attributes with more accurate attribute co-occurrences, but these methods still use heuristic or probabilistic structures learned from aggregated level data. By doing so, these approaches may work when regions are demographically similar, but when regions are demographically heterogeneous, using a fixed heuristic can misrepresent spatial non-stationarity of the synthetic population's attributes and introduce internal bias into the resulting synthetic population \cite{roxburgh2025,krupskii2018,mondal2024,bastin2023}. For example, a university region with a distinctive ``young + low-income + high-education'' attribute co-occurrence pattern does not share the same age--income--education co-occurrence pattern as a manufacturing region.

As for combinatorial optimization (CO), this method provides the capability to reconstruct region-specific joint distributions of a multi-attribute synthetic population dataset by preserving the spatial non-stationarity. The CO approach relies on individual-level data (e.g., PUMS microdata) that captures region-specific joint distribution of attributes, and uses it as an input to create a pool of candidate synthetic individuals. CO then iteratively duplicates or removes individuals from the pool until the resulting synthetic population matches target aggregate constraints (e.g., census tract–level data). Fitness evaluation and optimized sampling procedures are commonly used to assess and improve the quality of the generated synthetic population\cite{harland2012}. By doing so, the CO approach generates a synthetic population that preserves the region-specific joint distribution of attributes and provides an accurate representation of spatial non-stationarity. However, the synthesis process is usually computationally expensive and time-consuming, because each target region's synthesis process is a standalone process \cite{williamson1998}. Such a standalone process may limit the generalizability of CO, as they typically generate synthetic populations for a specific region and cannot be directly applied or transferred to other regions \cite{harland2012}. To overcome such limitations, generative models have provided a new possibility to generate a synthetic population across regions using a well-trained and tuned model \cite{sane2025,stoian2024}.

In addition, studies found that generative models such as Generative Adversarial Network (GAN), Variational Autoencoder (VAE) and diffusion model are suitable for the creation of synthetic populations, where the resulting population has reconstructed the joint distribution and preserved the spatial non-stationarity of the attributes \cite{goodfellow2014, kingma2014, xu2019ctgan, sane2025, ho2020}. But we did not use GAN and VAE, because such methods prefer to generate the most common attribute combinations in a specific region, which may result in losing the heterogeneities of the real population \cite{stoian2024, kotelnikov2023}. In the case of creating a five-attribute synthetic population within a certain area, such methods may ignore the rare attribute combinations. 

Recent works (e.g., \cite{tang2025, kang2023}) demonstrate that diffusion models can utilize individual-level data (e.g., PUMS) containing real region-specific joint distributions as a training target for model training. This allows the diffusion model to generate all possible attribute combinations using aggregated-level census data as conditions through a denoising process. As a result, using a diffusion model could overcome the common limitation of imposing fixed heuristics for attribute co-occurrences, which often leads to rare attribute combinations being ignored or underrepresented by the methods discussed above. In addition, utilizing a diffusion model could potentially improve the generalizability when generating a synthetic population across different geographic regions.

At the same time, location information of synthetic population has become an important feature to explore various topics within urban systems via geo-simulation methods (e.g., micro-simulation and agent-based modeling), such as human mobility, public health, and disaster resilience \cite{chapuis_gen_2018, yin_vac_2024, kim2020location}. To fulfill such demands, this work aims to create a multi-attribute geographically-explicit synthetic population that preserves both the joint distribution of multi-attributes and their spatial non-stationarity. To achieve this, we propose a generative framework based on a diffusion model. Specifically, the proposed framework utilizes open data (e.g., U.S. Census data) to create a five-attribute (e.g., age, gender, employment, education, income) synthetic population along with geographically-explicit location information for the whole United States of America (i.e., 50 States and Washington, D.C.).

In the following part of this paper, Section \ref{sec:method} introduces the details of the generative framework. Section \ref{sec:results} shows the resulting synthetic population along with the validation results. Lastly, Section \ref{sec:conclude} concludes this work and identifies the future works.  


\section{Methodology} \label{sec:method}

\subsection{Overview}
\begin{figure}[H]
\centering
\includegraphics[width=\linewidth,height=0.78\textheight,keepaspectratio]{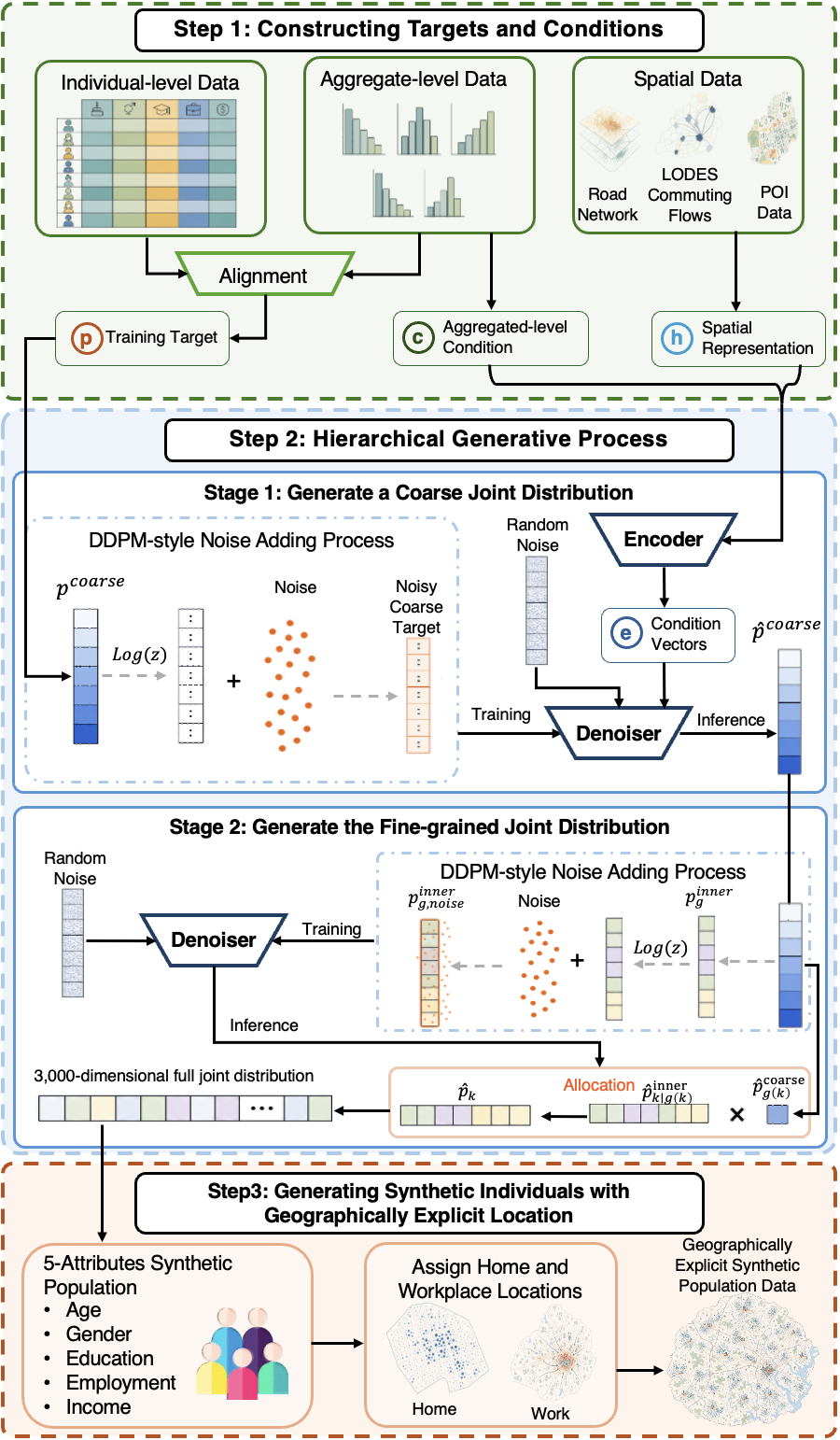}
\caption{Generative Framework.}
\label{fig:framework}
\end{figure}

As discussed in Section~\ref{sec:intro}, our goal is to create a geographically explicit synthetic population for the whole United States of America (50 States and Washington, D.C.) that can preserve the joint distribution and spatial non-stationarity of the five attributes. The resulting dataset contains 332,387,543 synthetic individuals across 2,462 PUMAs in the 50 states and Washington, D.C. Each individual has age, gender, employment, education, income, and latitude and longitude for the home location; employed individuals are additionally assigned latitude and longitude for the work location. Figure~\ref{fig:framework} demonstrates the framework based on a diffusion model and the following section describes this in detail. 

\subsection{Data Sources}\label{sec:data_sources}

Similar to all synthetic population work, data is also the key part for this work. As shown in Figure~\ref{fig:framework}, the proposed framework requires three types of data: individual-level data, aggregated-level census data and spatial data for spatial representation extraction and explicit location assignments. Table~\ref{tab:data} summarizes the corresponding sources and their roles in the framework.

For the individual-level data, we use the 2023 five-year American Community Survey (ACS) Public Use Microdata Sample (PUMS), which covers 2,462 Public Use Microdata Areas (PUMAs) \cite{census_puma}. Within each PUMA, ACS PUMS represents approximately 5\% of the population and reports individual-level survey responses on age, gender, income, education, and employment. These microdata allow us to estimate the region-specific joint distribution of the five attributes, which we use them as the training target for synthetic population generation \cite{acs_pums_handbook}.

For the aggregated census data, we use the ACS 5-year Detailed Tables, which provide official summary statistics for each PUMA \cite{acs, census_acs5_api}. This data can be retrieved using the Census API and used as the condition vector during both training and inference stages in Step~2 within the proposed framework. The rationale behind this is that such data can reflect statistical relationships of individual-level attributes from an aggregated perspective, which can be used as a training condition for the proposed framework. In addition, aggregated data are more accessible for researchers to acquire and use as the input to generate the synthetic population with the proposed framework.

Lastly, we use several spatial datasets, including road networks collected from OpenStreetMap \cite{openstreetmap}, point-of-interest (POI) data \cite{dataplor_poi} and LODES commuting flows \cite{lodes}. The POI data and LODES commuting flows are used to construct the condition vectors in Step~1. The rationale behind this is that these data provided spatial information beyond aggregated-level census data, allowing the model in Step~2 to learn from the spatial representation to preserve the spatial non-stationarity. The road-network data are used to assign explicit geographical locations for the synthetic population, and commuting-flow data provide the potential destination locations for the synthetic population's workplace assignments. With respect to the standardization, all the spatial data use the NAD83 geographic coordinate system (i.e., EPSG:4269).

\begin{table}[H]
\centering
\footnotesize
\caption{Data sources used in the main framework.}
\label{tab:data}
\begin{tabularx}{\textwidth}{p{0.27\textwidth}X p{0.29\textwidth}p{0.13\textwidth}}
\toprule
Data name & Data description & Role & Source \\
\midrule
Public Use Microdata Sample (PUMS) & Individual-level records for each Public Use Microdata Area (PUMA) & Construct the training target $\mathbf{p}$, defined as the joint distribution of the five variables for each PUMA & \cite{pums,acs_pums_handbook,census_puma} \\
American Community Survey (ACS) Detailed Tables & Aggregated-level census data for each PUMA and tract & Construct the condition vector $\mathbf{c}$ in Step~2 and allocate sampled residents across tracts in Step~3 & \cite{acs,census_acs5_api} \\
Point-of-interest (POI) data & POI data aggregated to PUMA level & Construct the spatial representation vector $\mathbf{h}$ used in Step~2 & \cite{dataplor_poi} \\
Longitudinal Employer--Household Dynamics Origin-Destination Employment Statistics (LODES) & Commuting-flow data derived from LODES & Construct the spatial representation condition vector $\mathbf{h}$ in Step~2 and provide destination tracts for workplace assignment in Step~3 & \cite{lodes} \\
Road network & Road networks collected from OpenStreetMap & Assign home and work coordinates within the assigned tracts in Step~3 & \cite{openstreetmap} \\
\bottomrule
\end{tabularx}
\end{table}

\subsection{Step 1: Create Target and Condition Vectors}
\label{sec:training_target}

As shown in Figure~\ref{fig:framework}, Step 1 prepares the training targets and the conditional vectors required for Step 2. To create the training target $\mathbf{p}$, individual-level PUMS data are utilized to extract a real region-specific joint distribution of the five variables within each Public Use Microdata Area (PUMA), which directly corresponds to the five target attributes of the synthetic population. Before the creation, we conduct the variable alignment for each PUMA. Specifically, the original numeric values of each variable in the PUMS dataset are mapped and converted into categorical bins using the variable definitions in the aggregate-level data (i.e., the American Community Survey 5-year Detailed Tables). For example, if an individual is 4 years old, then this individual's age is altered to the 0-5 age group, which is a variable name in the aggregated level data. Table~\ref{tab:var_attr} summarizes this variable alignment of the ACS and PUMS data. Then, a 3,000-dimensional vector is initialized, where each cell represents a potential five-variable combination. The target vector $\mathbf{p}$ is then created by mapping the empirical probabilities of these combinations using the individual-level PUMS data. A unique $\mathbf{p}$ is created for each PUMA, resulting in a total of 2,462 distinct training targets across the United States.

\begin{table}[H]
\centering
\caption{Variable Alignment Summary}\label{tab:var_attr}
\begin{tabular}{clc}
\toprule
\begin{tabular}[c]{@{}c@{}}Target \\ Attributes of\\ Synthetic\\  Population\end{tabular} & \begin{tabular}[c]{@{}l@{}}Variable \\ Names in\\ PUMS\end{tabular} & \begin{tabular}[c]{@{}c@{}}Variable \\ Names in\\ ACS\\ Detailed Tables\end{tabular} \\ \midrule
Age                                                                                       & AGEP                                                                & B01001                                                                               \\ 
Gender                                                                                    & Gender                                                              & B01001                                                                               \\ 
Education                                                                                 & SCHL                                                                & B15003                                                                               \\ 
Employment                                                                                & ESR                                                                 & B23025                                                                               \\ 
Income                                                                                    & PERNP                                                               & B20001                                                                               \\ \bottomrule
\end{tabular}

\end{table}

We then construct the conditional vectors, $\mathbf{c}$, using the aggregated level data by converting each categorical variable's absolute count into a probability distribution, resulting in a vector for a certain statistical area. As discussed in Section \ref{sec:data_sources}, aggregated level data is using PUMA as the statistical area, thus $\mathbf{c}$ reflects the real marginal distribution of the five variables of each PUMA, which can act as a condition during the synthesis process. Because $\mathbf{c}$ provides the variations of the population characteristic within a certain PUMA, spatial non-stationarity of the five variables can be preserved. 

Other than $\mathbf{c}$, the spatial representation condition vector $\mathbf {h}$ is created using the spatial data (i.e., POI and commuting-flow data) to preserve the spatial non-stationarity. For the creation of $\mathbf {h}$, the spatial data (i.e., POI and commuting-flow data) are aggregated to PUMA level and encoded with a vector value to differentiate each PUMA's functional characteristics. For example, if a PUMA has a high density of commercial POIs and a high volume of commuting inflows, this PUMA is encoded to reflect an employment or workplace region. The details of the $\mathbf {h}$ is provided in \ref{ssec:sptial_vector}. By doing so, $\mathbf {h}$ enables each PUMA to have a unique spatial representation and provides the potential for this generative framework to preserve the non-stationarity during synthesis. At the end of Step 1, across the United States, there are 2,462 vector sets $(\mathbf{p}, \mathbf{c}, \mathbf{h})$ for each PUMA, which will be the inputs for Step~2.



\subsection{Step 2: Hierarchical Generative Process}
\label{sec:training_model}

As noted in Section \ref{sec:intro}, Step 2 serves as the core of this generative framework. It is built upon a diffusion model that learns from the vector sets $(\mathbf{p}, \mathbf{c}, \mathbf{h})$ constructed in Step 1 to predict the joint distribution of the five variables within the training target vector for each PUMA. The framework utilizes a hierarchical structure, employing the diffusion model twice to generate the joint distribution under different conditioning vectors for each PUMA. The rationale for this design is that reconstructing a five-variable joint distribution may result in too many variable combinations to train and predict stably within a one-stage diffusion process \cite{zhang21z, qiang23a}.

\subsubsection{Stage 1: Generate a Coarse Joint Distribution}

Specifically, Stage~1 first generates a coarse joint distribution, denoted as $\hat{\mathbf{p}}^{\mathrm{coarse}}$ using individual-level target $\mathbf{p}$, aggregated-level condition $\mathbf{c}$ and spatial representation $\mathbf{h}$. Firstly, $\mathbf{c}$ and $\mathbf{h}$ are concatenated and passed through a multilayer perceptron (MLP) condition encoder to produce a learned condition vector $\mathbf{e}$, because they come from different data sources and have different dimensions. This learned condition vector $\mathbf{e}$ is used by the Stage~1 denoiser to predict coarse probability values $\hat{\mathbf{p}}^{\mathrm{coarse}}$.

Then, the individual-level target $\mathbf{p}$ is converted to a coarse version, denoted as $\mathbf{p}^{\mathrm{coarse}}$. As noted in Step 1, the $\mathbf{p}$ is created based on categorical data using the variable names from American Community Survey 5-year Detailed Tables, which is further grouped into broader categories, summarized in Table~\ref{tab:coarse_target}, where gender is kept as binary (i.e., 1 for male and 2 for female). Considering the computational efforts to train the diffusion model and the resulting synthetic population's fidelity, we construct the coarse target $\mathbf{p}^{\mathrm{coarse}}$ to reduce the computational burden of Stage~1. Since Stage~2 refines each coarse variable combination into its fine-grained combinations, the grouped categories need to correspond to demographically meaningful parent categories. For example, an age group of 18--34 can be refined into 18--24 and 25--34, so the Stage~1 prediction remains interpretable as the probability mass assigned to young adults. 


Therefore, the category reduction is defined by grouping adjacent or substantively similar categories from the original variables. This keeps the coarse target compact while preserving the demographic meaning needed for the fidelity of the final synthetic population. The categories of each variable are reduced as follows: Age is compressed from 10 fine categories to 4 coarse groups; Gender remains unchanged as a binary variable; Education remains at its original 5-category resolution; Employment is consolidated from 5 categories to 4 by grouping unemployed and armed-forces categories; and Income remains at its original 6-category resolution. As a result, $\mathbf{p}^{\mathrm{coarse}}$ has 960 coarse combinations ($4 \times 2 \times 5 \times 4 \times 6$), compared with 3,000 fine-grained combinations in $\mathbf{p}$ ($10 \times 2 \times 5 \times 5 \times 6$). This $K=960$ setting is selected based on a sensitivity analysis (See Supplementary Figure~\ref{fig:s_k_sweep} in Section \ref{ssec:sensitivity}).


\begin{table}[H]
\centering
\small
\caption{Construction of the Stage~1 coarse target.}
\label{tab:coarse_target}
\begin{tabularx}{\textwidth}{lccX}
\toprule
Variable & Fine categories & Coarse categories & Coarse grouping \\
\midrule
Age & 10 & 4 & $0$--$17$; $18$--$34$; $35$--$64$; $65+$ \\
Gender & 2 & 2 & Unchanged \\
Education & 5 & 5 & Unchanged \\
Employment & 5 & 4 & Under 16; employed; unemployed or armed forces; not in labor force \\
Income & 6 & 6 & Unchanged \\
\midrule
Joint target & $10 \times 2 \times 5 \times 5 \times 6$ & $4 \times 2 \times 5 \times 4 \times 6$ & $3{,}000$ fine combinations are reduced to $960$ coarse combinations \\
\bottomrule
\end{tabularx}
\end{table}

The coarse target $\mathbf{p}^{\mathrm{coarse}}$ is then transformed into standardized log-probability values, which are used to create a noisy coarse target through a known noise adding process. Specifically, we add Gaussian noise with different noise levels using a discrete-time Gaussian denoising diffusion probabilistic model (DDPM)-style implementation \cite{ho2020,kotelnikov2023}. 

The diffusion model in Stage 1 is trained using the known noise adding process along with aggregated-level condition $\mathbf{c}$ and the spatial representation $\mathbf{h}$. Consequently, this model becomes a denoiser that predicts the probability distribution of coarse variable combinations within a given PUMA while preserving spatial non-stationarity. Once trained, the model can be further utilized to infer the coarse joint distribution $\hat{\mathbf{p}}^{\mathrm{coarse}}$ using the inputs of noisy coarse target and condition vectors(i.e., $\mathbf{c}$ and $\mathbf{h}$). 

At the end of this stage, one $\hat{\mathbf{p}}^{\mathrm{coarse}}$ is generated for each PUMA. Within $\hat{\mathbf{p}}^{\mathrm{coarse}}$, there are 960 coarse variable combinations and their probability values.

\subsubsection{Stage 2: Generate the Fine-grained Joint Distribution}

Stage~2 utilizes another diffusion model along with  $\hat{\mathbf{p}}^{\mathrm{coarse}}$ to generate the probability values of potential fine-grained variable combinations $\hat{\mathbf{p}}$, which reflect a full five-variable joint distribution. To generate this, a $g$ is extracted from$\hat{\mathbf{p}}^{\mathrm{coarse}}$, which represents one coarse group (i.e., coarse variable combinations) and maps each coarse group back to its fine-grained variable combinations. Specifically, $g$ guides the creation of the inner-coarse refinement target $\mathbf{p}^{\mathrm{inner}}_{g}$, which 
allocates $\mathbf{p}$'s joint probability values to the fine-grained variable combinations within a certain coarse group. 

Before training the diffusion model, each probability value within $\mathbf{p}^{\mathrm{inner}}_{g}$ is normalized to a sum of 1 ($\sum \mathbf{p}^{\mathrm{inner}}_{g} = 1$) based on its corresponding coarse group $g$, resulting in a total of 960 normalized vectors $\mathbf{p}^{\mathrm{inner}}_{g}$. As for the training process, each normalized inner-coarse refinement target $\mathbf{p}^{\mathrm{inner}}_{g}$ is transformed into standardized log-probability values and noised using the same DDPM-style noise adding strategy as Stage~1, which is denoted as $\mathbf{p}^{\mathrm{inner}}_{g, \mathrm{noise}}$. Then, this stage's diffusion model is trained with this known noise adding process along with coarse group $g$ and its coarse probability value. By doing so, this diffusion model can be a denoiser to predict the known noise adding process from the noisy inner-coarse refinement target (i.e, $\mathbf{p}^{\mathrm{inner}}_{g, \mathrm{noise}}$). Consequently, during the inference process, the trained diffusion model can reverse the noise to generate the inner-coarse refinement probability values (i.e., $\hat{\mathbf{p}}^{\mathrm{inner}}_{g}$) using the coarse joint distribution $\hat{\mathbf{p}}^{\mathrm{coarse}}$ from Stage 1. These inner-coarse probability values guide the probability values' allocation for fine-grained variable combinations.

To allocate the final probability values for the fine-grained variable combinations, it assigns the final joint probability value to each fine-grained combination by multiplying the predicted coarse probability values from Stage 1 with the corresponding probability values within inner-coarse refinement probability values (i.e., $\hat{\mathbf{p}}^{\mathrm{inner}}_{g}$) generated in Stage 2. For example, consider one coarse group presents a variable combination with age in the 18--34 category, gender in category 1, education in the high-school-or-GED category, employment in the unemployed-or-armed-forces category, and income in the \$50k--\$75k category. This coarse group contains four fine-grained variable combinations. The age category can be ungrouped into 18--24 and 25--34, the employment category can be ungrouped into unemployed and armed-forces categories. Gender, education, and income remain unchanged.

The final joint probability assignment process can be written as Equation~\ref{eq:ungroup}. Let $k$ denote a fine-grained variable combination, and let $g(k)$ denote the coarse variable combination that contains $k$. The final predicted probability value for $k$ is defined as
\begin{equation}
  \hat{p}_k = \hat{p}^{\mathrm{coarse}}_{g(k)} \hat{p}^{\mathrm{inner}}_{k \mid g(k)} ,
  \label{eq:ungroup}
\end{equation}

where $\hat{p}^{\mathrm{coarse}}_{g(k)}$ is the coarse probability weight predicted by Stage~1, and $\hat{p}^{\mathrm{inner}}_{k \mid g(k)}$ is the within-coarse probability value predicted by Stage~2.

This stages convert the 960-dimensional coarse joint distribution $\hat{\mathbf{p}}^{\mathrm{coarse}}$ back to the 3,000-dimensional full joint distribution $\hat{\mathbf{p}}$. By the end of this stage, each PUMA has a 3,000-dimensional full joint distribution over the five variables, which are further used to sample the synthetic individuals in Step 3. 

\begin{figure}[H]
\centering
\includegraphics[width=\textwidth]{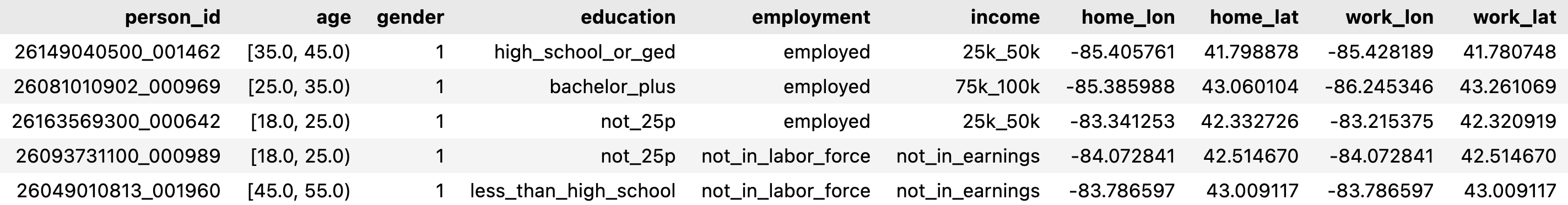}
\caption{Sample of Resulting Synthetic Population}
\label{fig:sample_data}
\end{figure}

\subsection{Step 3: Generate Synthetic Individuals with Geographically Explicit Locations}
\label{sec:inference}

As shown in Figure~\ref{fig:framework}, Step~3 generates a geographically explicit synthetic population by turning the predicted joint distribution $\hat{\mathbf{p}}$ from Step~2 into synthetic individuals with five attributes and explicit home locations along with work locations, if applicable.

As noted in Step 2, each PUMA's 3,000-dimensional full joint distribution over the five variables $\hat{\mathbf{p}}$, where each probability value indicates the likelihood of drawing a specific five-attribute combination. To generate the synthetic individuals, we perform independent random sampling based on these joint probabilities. For example, if a given PUMA $r$ requires a population of size $N_r$, we execute $N_r$ independent draws, where the specific five-attribute combination assigned to each synthetic individual is driven directly by the corresponding probability values within $\hat{\mathbf{p}}$. Consequently, each sampled individual receives a coherent, co-occurring set of age, gender, education, employment, and income attributes. 

After generating the synthetic individuals, we first assign the home geographically explicit location, then assign the workplace location. As for the assignment of the home location, we use the census tract boundary as the inner-PUMA allocation unit. Because one of the aggregated-level data (i.e., American Community Survey 5-year Detailed Tables) provides the number of population and their attribute at the tract-level, which we can use to guide the allocation of synthetic individuals to specific tracts. Specifically, this ACS tract-level data is used to initialize a matrix ${N}_{t,k}$. It contains the potential number of synthetic individuals with a certain five-attribute combination $k$ that will be assigned to a specific tract $t$, where each cell represents how many individuals of the five-attribute combination $k$ are assigned to the tract $t$. 

To build the matrix ${N}_{t,k}$, we extract the values of the variables in ACS tract-level data that correspond to the five-attribute using Equation \ref{eq:tract_allocation_initialization}. 

\begin{equation}
\begin{aligned}
    s_{t,a,v} &= \frac{N_{t,a,v}}{\sum_{v'}N_{t,a,v'}}, \\
    q(t\mid k) &= \frac{\prod_{a\in A}s_{t,a,k_a}}{\sum_{t'\in r}\prod_{a\in A}s_{t',a,k_a}}, \\
    N_{t,k} &\leftarrow N_{r,k}q(t\mid k),
\end{aligned}
\label{eq:tract_allocation_initialization}
\end{equation}

where $N_{t,a,v}$ represents the total number of individuals with attribute $a$ and its value is $v$ in a certain census tract $t$; $A$ is five attributes; $k_a$ is one of the attributes in combination $k$; and $N_{r,k}$ is the number of synthetic individuals with a certain attribute combination $k$ in a PUMA $r$. The denominator in $q(t\mid k)$ sums over all tracts within a PUMA $r$, so $q(t\mid k)$ gives the initial tract-assignment probability for combination $k$.

Then, the values within matrix $N_{t,k}$ are iteratively adjusted through the proportional fitting process to satisfy two constraints (e.g., joint distribution and marginal distribution). As for the joint-distribution constraint, it provides the constraint for the allocation of synthetic individuals in a PUMA $r$ with attribute combination $k$ to a certain tract $t$. This means the sum of the joint distributions across every tract within a given PUMA must equal the predicted 3,000-dimensional full joint distribution $\hat{\mathbf{p}}$. To adjust values based on this constraint, if a PUMA has 100 synthetic individuals with attribute combination $k$ but is currently allocating only 80 based on the sum of $N_{t,k}$, then the cells of column $k$ within matrix $N_{t,k}$ are multiplied by $100/80$. This ensures that the aggregated joint distribution of all tracts follows the PUMA-level target.

As for the marginal distribution constraint, it provides the constraint for specific attributes of the synthetic individuals. We select age and gender for this constraint because the records of these two attributes in the ACS data follow a known joint distribution. By using age and gender as constraints, it could potentially reduce the randomness of the allocation, ensuring that the synthetic population's age-gender joint distributions remain representative comparing to the real population. To utilize this constraint, the total number of synthetic individuals within a specific age and gender group in a given tract $t$ must match the corresponding census record $N_{t,a,v}$. To adjust the values using this constraint, for example, if tract $t$ currently contains 160 male synthetic individuals within the 18–34 age group while the actual census records 200, then all cells in tract $t$ whose attribute combinations $k$ include males aged 18–34 within the matrix $N_{t,k}$ are multiplied by $200/160$.

To allocate synthetic individuals to their home tracts, the values within the allocation matrix $N_{t,k}$ are iteratively adjusted until they satisfy both constraints. Specifically, the iterative adjustment process compares the values within $N_{t,k}$ with the corresponding census records $N_{t,a,v}$ by applying a strict convergence threshold with a numeric tolerance of $10^{-6}$. As a result, the adjusted matrix $N_{t,k}$ is utilized to guide the allocation of synthetic individuals with a specific five-attribute combination $k$ to a certain tract $t_h$. After the allocation of tract $t_h$, geographically-explicit home location (i.e., latitude and longitude) is assigned to each synthetic individual, where the home locations are created approximately 50 meters apart along the residential road network. The rationale behind this design is to avoid potential privacy issues \cite{jiang_large-scale_2024}.

With respect to the assignment of workplace locations, we firstly extract the synthetic individuals with valid employment information (i.e., have a job), then assign their work tracts using LODES commuting flows. The synthetic individuals' home tract $t_h$ is utilized to find a work tract $t_w$ from LODES based on the probability $P(t_w \mid t_h)$ calculated using \ref{eq:work_tract_probability}. For example, if an origin tract $t_h$ has 10 potential destination tracts, the destination tract with a higher inflow $F$ has a higher probability of being selected as the destination work tract $t_w$. After having the work tract $t_w$, a geographically-explicit workplace location (i.e., latitude and longitude) is assigned to each synthetic individual. For the creation of workplace locations, we use the method from previous work \cite{jiang_large-scale_2024}, where we create the spaces on secondary roads 20 meters apart and the intersections of residential roads within the road network. 

\begin{equation}
P(t_w \mid t_h)=\frac{F_{t_h,t_w}}{\sum_{t'_w} F_{t_h,t'_w}},
\label{eq:work_tract_probability}
\end{equation}
where $F_{t_h,t_w}$ denotes the commuting flow count from home tract $t_h$ to work tract $t_w$. By doing so, we assign a potential work tract $t_w$ to each sampled worker. 

At the end of this step, a five-attribute geographically explicit synthetic population is created. Each individual has age, gender, employment, education, income, and latitude and longitude for the home location; employed individuals are additionally assigned latitude and longitude for the work location.

\section{Results} \label{sec:results}

\subsection{Resulting Geographically Explicit Synthetic Population Datasets}
\label{sec:data_reporting}

The framework generates a geographically explicit synthetic population for the whole United States of America (i.e., 50 states and Washington, D.C.). The resulting dataset contains 332,387,543 synthetic individuals across 2,462 PUMAs. Each synthetic individual has age, gender, education, employment, income, along with home and workplace location latitude and longitude, if applicable. The resulting datasets are stored using .csv format and organized by state, which has been made available on OSF \url{https://osf.io/e7wp8/} for researchers who see fit. Figure \ref{fig:sample_data} shows the resulting population dataset using a sample and Table \ref{tab:sample_attributes} shows the synthetic population's attributes and their descriptions.  
\begin{table}[H]
\centering
\small
\caption{Attributes in the synthetic population sample.}
\label{tab:sample_attributes}
\begin{tabular}{p{0.15\textwidth} p{0.28\textwidth} p{0.50\textwidth}}
\toprule
Attribute & Description & Values \\
\midrule

person\_id & Individual unique identifier & String: tract GEOID followed by an underscore and a zero-padded within-tract sequence number, e.g., 26133970600\_001938. \\

age & Age group & \begin{tabular}[t]{@{}l@{}}
{}[0.0, 5.0), [5.0, 18.0), [18.0, 25.0) \\
{}[25.0, 35.0), [35.0, 45.0), [45.0, 55.0) \\
{}[55.0, 65.0), [65.0, 75.0), [75.0, 85.0) \\
{}[85.0, 1000.0)
\end{tabular} \\

gender & Gender & Integer: 1 male; 2 female. \\

education & Educational attainment group & \begin{tabular}[t]{@{}l@{}}
not\_25p: under age 25 \\
less\_than\_high\_school \\
high\_school\_or\_ged \\
some\_college\_or\_assoc \\
bachelor\_plus
\end{tabular} \\

employment & Labor force status & \begin{tabular}[t]{@{}l@{}} not\_16p: under age 16 \\ employed \\ unemployed \\ armed\_forces \\ not\_in\_labor\_force \end{tabular} \\

income & Individual earnings group & \begin{tabular}[t]{@{}l@{}}
not\_in\_earnings \\
lt\_25k \\
25k\_50k \\
50k\_75k \\
75k\_100k \\
ge\_100k
\end{tabular} \\

home\_lon & Longitude of home location & Float in NAD83 (EPSG:4269) coordinates, e.g., -85.502835. \\

home\_lat & Latitude of home location & Float in NAD83 (EPSG:4269) coordinates, e.g., 43.905819. \\

work\_lon & Longitude of workplace location & Float in NAD83 (EPSG:4269) coordinates, e.g., -84.402919. \\

work\_lat & Latitude of workplace location & Float in NAD83 (EPSG:4269) coordinates, e.g., 44.802813. \\

\bottomrule
\end{tabular}
\end{table}

\subsection{Validation}
\label{sec:technical_validation} 

In this work, we conducted two rounds of validation experiments, including an internal validation to confirm that the resulting datasets match the census data and a held-out regional experiment to assess the framework's generalizability.

\begin{figure}[H]
\includegraphics[width= \linewidth]{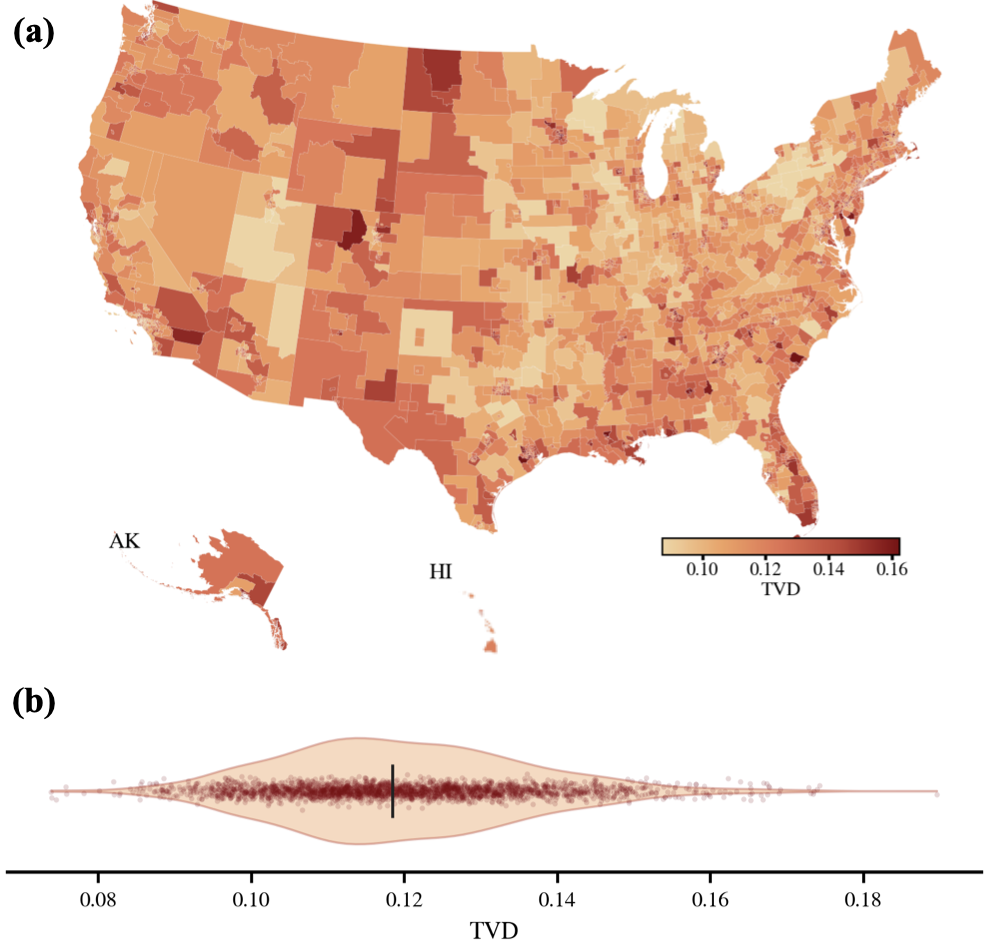}
\caption{Validation Result of Joint Distribution in PUMA-level: (a) Puma-level TVD difference; (b) Overall distribution TVD.}
\label{fig:national_puma_tvd}
\end{figure}

As for the internal validation, we use two input datasets as the benchmark, which are aggregated-level census data at PUMA-level along with the PUMS data (discussed in Section \ref{sec:data_sources}). As discussed in Section \ref{sec:intro}, the goal of this work is to reconstruct the joint distribution of the synthetic population's five attributes. Thus, we compare the all resulting synthetic population with the real PUMS data. To validate, total variation distance (TVD) has been used to measure the joint distribution \cite{lovelace2015}. Figure (a) illustrates the TVD's difference between the $\mathbf{p}$ and  $\hat{\mathbf{p}}$, in which darker orange indicates larger TVD. As shown in Figure \ref{fig:national_puma_tvd} (b), we compare the TVD in the training target $\mathbf{p}$ and the predicted 3,000-dimensional full joint distribution $\hat{\mathbf{p}}$, the mean TVD is 0.116, which represents 88.4\% overlap between the predicted joint distribution and the target joint distribution. In addition, both $\mathbf{p}$ and $\hat{\mathbf{p}}$ represent each PUMA's region-specific joint distributions. As a result, the framework could reconstruct region-specific joint distributions across all PUMAs that are demographically heterogeneous. To ensure the robustness of this frame, we conduct multi-seed experiments. The mean TVD of the experiments is 0.11651 (SD = 0.01586), with a 95\% empirical interval of [0.08949, 0.14981].

\begin{figure}[H]
\centering
\includegraphics[width=\textwidth,keepaspectratio]{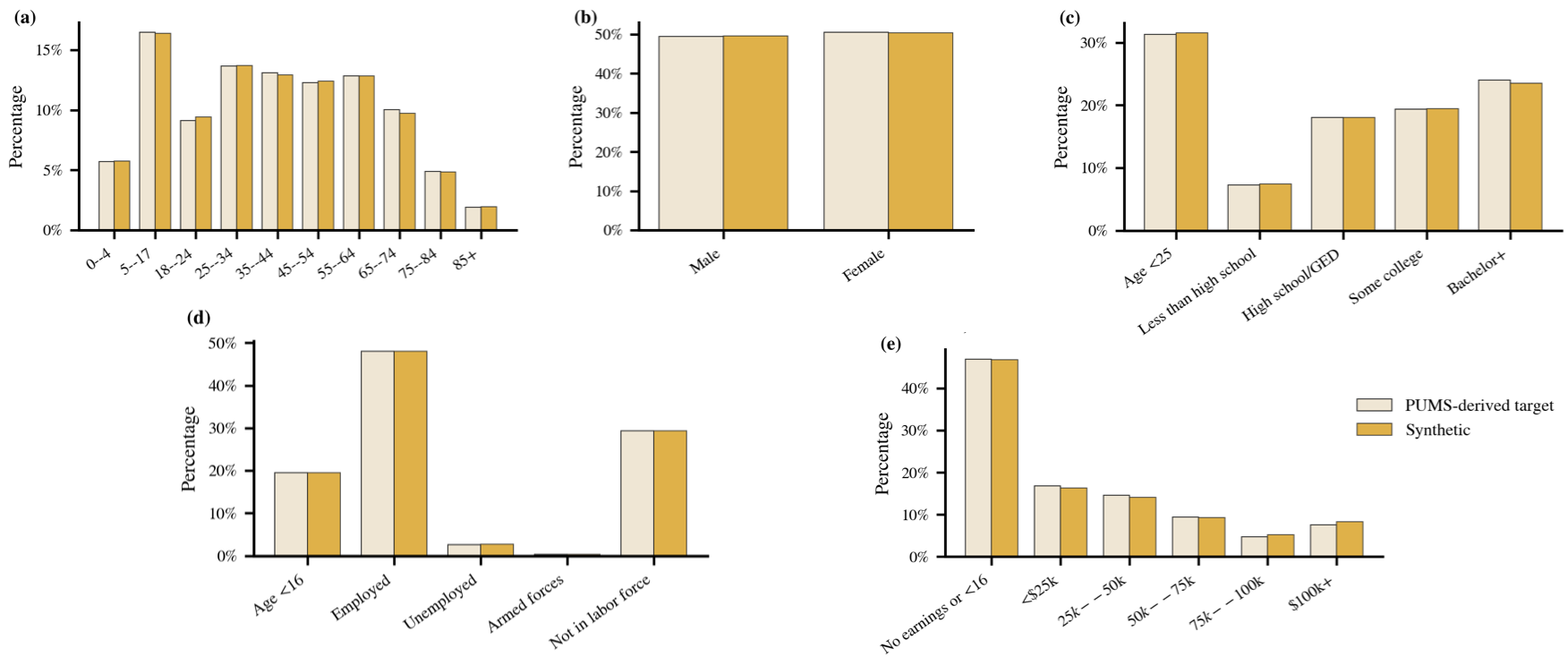}
\caption{Percentage of Population in Each Variable Group (Synthetic Population's Attributes versus Aggregated-level Marginal Distribution): (a) Age Group; (b) Gender; (c) Education; (d) Employment; (e) Income.}
\label{fig:national_attribute_distribution}
\end{figure}

\begin{figure}[H]
\centering
\includegraphics[width=\linewidth]{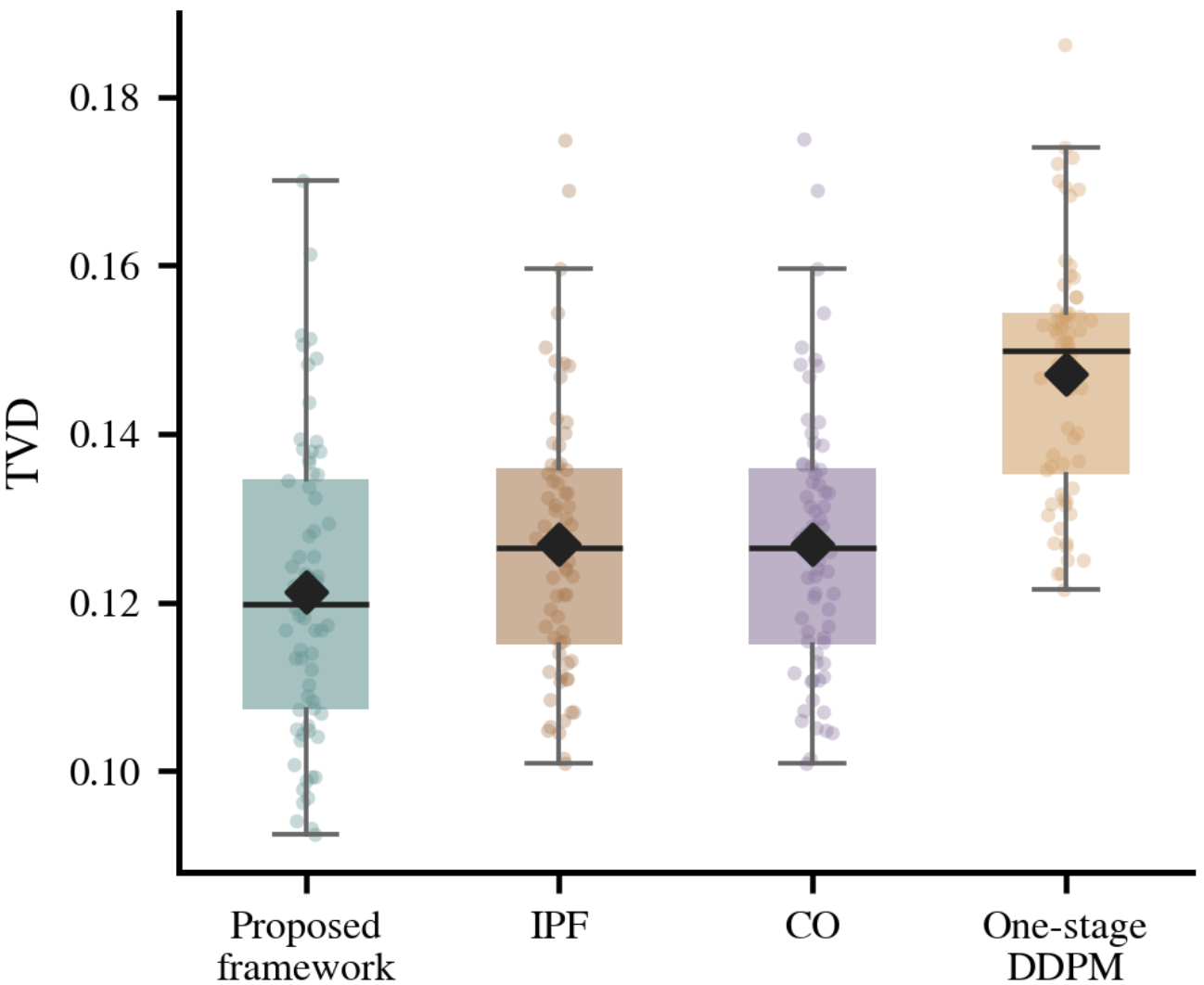}
\caption{TVD distributions across baseline methods.}
\label{fig:michigan_method_comparison}
\end{figure}

\begin{figure}[H]
\centering
\includegraphics[width=\textwidth,height=0.50\textheight,keepaspectratio]{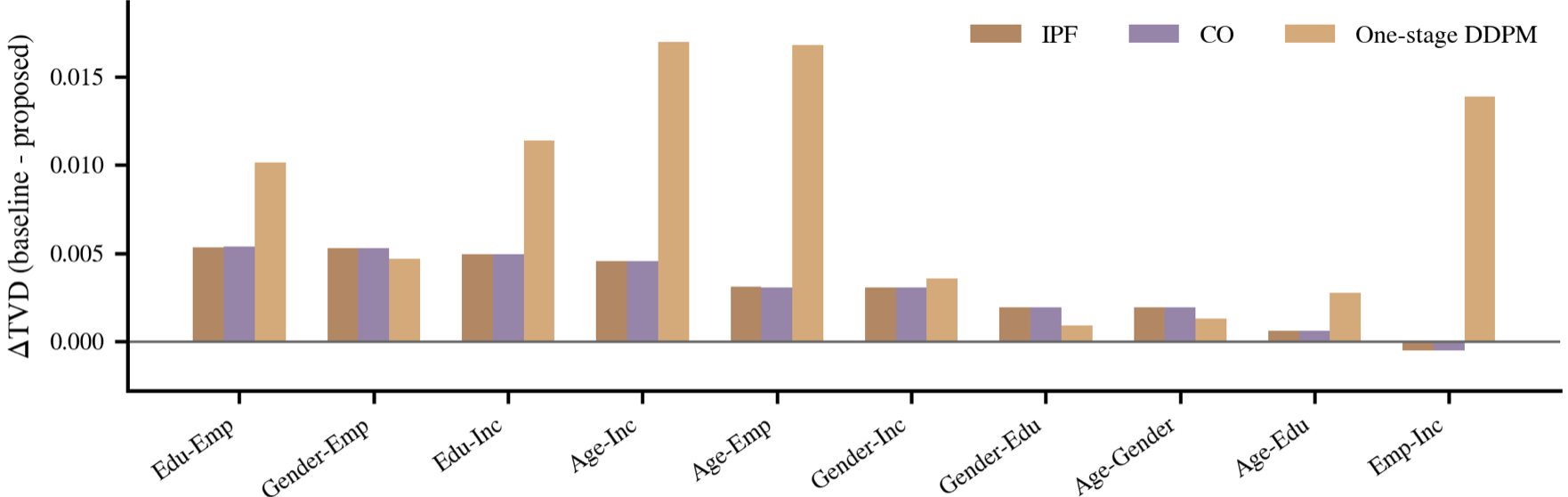}
\caption{The difference of TVD for each two-attribute between the baseline models and the proposed framework.}
\label{fig:pairwise_decomposition}
\end{figure}

Other than that, we further compare the marginal distributions of the synthetic population with the aggregated-level input data. Figure~\ref{fig:national_attribute_distribution} shows that the synthetic population preserves the marginal distribution of each attribute after individuals are sampled from the predicted joint distribution. We calculated the average absolute difference (AAD) between the synthetic population and the source data for each attribute. The resulting discrepancies are low, such as, 0.0012 for age, 0.0009 for gender, 0.0022 for education, 0.0002 for employment, and 0.0037 for income.

In addition, we conduct an experiment on a held-out state to validate the generalizability of the proposed generative framework, ensuring that the framework can generate a synthetic population for a region excluded from the training process. Simultaneously, this experiment confirms that the region-specific joint distributions of the five attributes within this excluded area are reconstructed. Thus, we exclude the state of Michigan as the held-out experiment state since it contains a heterogeneous set of urban and suburban regions, where all PUMAs remain excluded from model training. Other than Michigan, states held-out experiments are provided in Supplementary (Table~\ref{tab:S3_state_holdout_checks}). 

Within this experiment, we compare the proposed generative framework with three baselines within the synthetic population research domain: Iterative Proportional Fitting (IPF), combinatorial optimization (CO), and a one-stage Denoising Diffusion Probabilistic Model (DDPM) using TVD as the metric. IPF uses the target marginals derived from the aggregated-level census data and adjusts a fixed seed table, which represents the average joint distribution over training PUMAs. CO constructs a synthetic population by duplicating and deleting individuals from the available individual-level data pool. The one-stage DDPM baseline uses the same training target and condition vectors as the proposed framework, and predicts the full joint distribution directly within one stage. 

As shown in Figure \ref{fig:michigan_method_comparison}, the proposed generative framework generates the lowest TVD among the comparisons. Its mean TVD is $0.119$, compared with $0.12698$ for IPF, $0.12697$ for CO, and $0.14707$ for the one-stage DDPM baseline. These values correspond to relative TVD reductions of $6.3\%$ over IPF, $6.3\%$ over CO, and $19.1\%$ over the one-stage DDPM baseline.



To further ensure generalizability, we decompose the five-attribute joint distribution into two-attribute distributions to examine where the framework improves the reconstruction of attribute co-occurrences using the held-out regional experiment. Figure~\ref{fig:pairwise_decomposition} shows the results for each two-variable distribution and reports the differences in TVD between the baseline models and our proposed generative framework. The average difference is 0.00302 for IPF and 0.00302 for CO, while the one-stage DDPM gap is 0.00824, which means these baseline models underperform in reconstructing the joint distribution compared to the proposed framework. Especially for IPF and CO, the TVD difference for the age–gender attribute combination is smaller because the aggregate-level data directly provides the age and gender joint distribution. However, larger differences in TVD occur in the attribute combinations of education–employment, gender–employment, education–income, and age–income. Through this decomposition experiment, the framework has been shown to improve the reconstruction of attribute combinations whose joint distributions are not recorded in the aggregate census data (e.g., education and income), indicating its capability to infer latent socioeconomic dependencies.

As such, the internal validation shows the framework's capability to create a five-attribute geographically-explicit synthetic population for the whole United States that matches the individual-level and aggregate-level source data. Additionally, the held-out TVD comparison experiment and pairwise decomposition show that the generative framework reconstructs the region-specific full joint distribution for held-out regions, which further proves its generalizability to scale this geographically-explicit synthetic population across the entire nation.


\section{Conclusion and Discussion}\label{sec:conclude}

As discussed in Section~\ref{sec:intro}, we propose a generative framework based on a diffusion model to generate a geographically explicit synthetic population that can reconstruct region-specific joint distributions of multiple attributes and also preserve spatial non-stationarity. This model is trained on region-specific joint distributions derived from individual-level data, which enable the framework to generate a five-attribute geographically explicit synthetic population for the entire United States (including all 50 states and Washington, D.C.) using aggregate-level data as conditioning input. To validate the reconstruction of region-specific joint distributions, we calculate the TVD between the generated full joint distribution and the training target joint distribution across all PUMAs. The low mean TVD indicates that the proposed framework can generate PUMA-specific joint distributions while preserving spatial non-stationarity across regions. We further calculate the AAD between synthetic population's each attribute and the corresponding aggregate-level marginal variable, and the small discrepancies show that the generated population preserves the marginal structure of the condition input data. In addition to this internal validation, the held-out regional experiment with Michigan proves the framework's generalizability in a region excluded from the training process along with the pairwise decomposition experiment. Through these experiments, the framework not only shows the improvements of the inference for latent attribute dependencies comparing to baseline models, but also proves the framework's capability for generating a geographically explicit synthetic population across the whole United States.

Similar to all synthetic population work, there are always limitations to improve. First, using PUMA-level census data as condition input generates the synthetic population with an age group instead of an exact age, especially the age group 5-17, where the range is large. Thus, one direction of the future work would be to incorporate richer aggregate-level constraints to improve age synthesis. Second, daytime locations (e.g., educational sites) for individuals under the age of 18 are currently missing from the synthetic population, because the LODES data only provide the origin and destination census tract of adults who are employed. This could constrain the exploration of daily activity patterns from the mobility perspective, which could be further addressed by assigning school-age individuals to an education site extracted from POI data \cite{dataplor_poi}. Third, the location-assignment for homes and workplaces step is limited by the spatial data available. Specifically, as workplace locations depend on employment status, commuting flows, destination tracts, and within-tract placement, future work can incorporate finer-grained POI, land-use, employment, and mobility signals to strengthen the spatial assignment process. 

Even with these limitations, the proposed generative framework creates a five-attribute synthetic population for the whole United States with geographically explicit home and workplace locations. In addition, the region-specific joint distributions of the five attributes have been validated, which are reconstructed with high fidelity. Thus, this work provides a scalable and generalized framework to create a multi-attribute geographically-explicit synthetic population for both regional and national levels. By reconstructing region-specific joint distributions of these five attributes using this framework, the resulting synthetic population could introduce more realistic behaviors into geo-simulations, such as agent-based modeling, enabling further exploration of the emergence of complex urban phenomena through human interactions.

\section{Data Availability}
The resulting datasets are stored using .csv format and organized by state, which has been made available on OSF: \url{https://osf.io/e7wp8} 

\section{Code Availability}
All scripts for data processing, model training, sampling, and validation are available on GitHub \url{https://github.com/wujlin/Synthetic_City}.

\section*{Acknowledgments}
Guangdong Provincial Talent Program 2025D03J0019

\bibliographystyle{unsrtnat}
\bibliography{references}

\appendix
\setcounter{figure}{0}
\renewcommand{\thefigure}{S\arabic{figure}}
\renewcommand{\theHfigure}{S\arabic{figure}}
\setcounter{table}{0}
\renewcommand{\thetable}{S\arabic{table}}
\renewcommand{\theHtable}{S\arabic{table}}

\section{Supplementary Information}



\subsection{Construction of Spatial Representation Vector} \label{ssec:sptial_vector}

The spatial representation condition vector $\mathbf{h}$ describes PUMA-level functional characteristics using POI and commuting-flow data. The POI-related components describe local business activity, while the commuting-flow-related components describe how people commute between the home tract and the workplace tract. Both components are aggregated to the PUMA level and concatenated as additional condition information for Step~2.

Rationale of using POI (i.e., point of interest) data is that it describes the locations of businesses, services, institutions, and other activity places, which reflect the degree of business activity in different spaces \cite{dataplor_poi}. To construct the POI-related components, we organize these components into five levels. The first level encompasses the total POI count, log-transformed POI count, POI count density per $\mathrm{km}^2$, and log-transformed POI density. The second level describes category diversity and concentration, including the total number of unique POI categories, the POI percentages across categories, and the percentage share of the single largest POI category. Utilizing the POI categorization from Dataplor \cite{dataplor_poi}, the third level describes 14 broad activity categories using their respective counts and percentages. The fourth level captures 31 POI subcategories, each comprising its log-transformed counts and percentages. Finally, the fifth level details the percentages of POIs across 120 specific categories representing frequent business types. These five levels produce 218 POI-related components.

Rationale of using Commuting-flow data is that it represents each PUMA's employment connectivity, which describes how people commute between home and workplace regions. We create 17 commuting-flow-based features from LODES and organize them into six levels. The first level describes commuting counts and direction, including the number of commuters leaving the PUMA for work, the number of commuters entering the PUMA for work, and the difference between these two values on the log scale. The second level describes within-PUMA commuting using the percentage of commuters who both live and work within the same PUMA. The third level describes the distribution of commuting flows across connected PUMAs, including the distribution of outgoing commuters across workplace PUMAs, the distribution of incoming commuters across home PUMAs, and the distribution of overall commuting connections across other PUMAs. The fourth level describes top-flow percentages. These percentages measure the percentage of outgoing commuters traveling to the largest one or three workplace PUMAs, and the percentage of incoming commuters coming from the largest one or three home PUMAs. For example, a high top-flow percentage means that most commuters are connected to one or a few other PUMAs, while a low top-flow percentage means that commuting flows are distributed across many PUMAs. The fifth level records whether valid LODES commuting records are available for the PUMA. The sixth level describes within-state ranks for selected commuting measurements, including outgoing commuting counts, incoming commuting counts, within-PUMA commuting percentage, the spread of outgoing flows, and the distribution of incoming flows. Overall, these 17 measurements distinguish PUMAs that supply, attract, and retain commuters or have most commuting flows connected to only a few other PUMAs.

Finally, the 218 POI-based features and 17 commuting-flow-based features are concatenated to construct a 235-dimensional spatial representation condition vector $\mathbf{h}$.

\subsection{Sensitivity to the Coarse Variable Combinations Size}\label{ssec:sensitivity}
We further examine whether the Stage~1 coarse combination size affects held-out joint-distribution recovery. Under the same POI+LODES condition setting, we evaluate candidate values of $K$ using the held-out TVD metric across the 68 Michigan PUMAs. As shown in Supplementary Figure~\ref{fig:s_k_sweep}, the mean TVD is clearly higher when $K=288$ (0.123495), but the curve enters a stable plateau once $K$ reaches 720. Although $K=1800$ gives the lowest mean TVD (0.118709), its improvement over $K=960$ (0.118809) is only 0.00010. We therefore use $K=960$ as the default coarse state-space size, since it preserves plateau-level accuracy while keeping the Stage~1 prediction target more compact.

\begin{figure}[H]
\centering
\includegraphics[width=0.95\columnwidth]{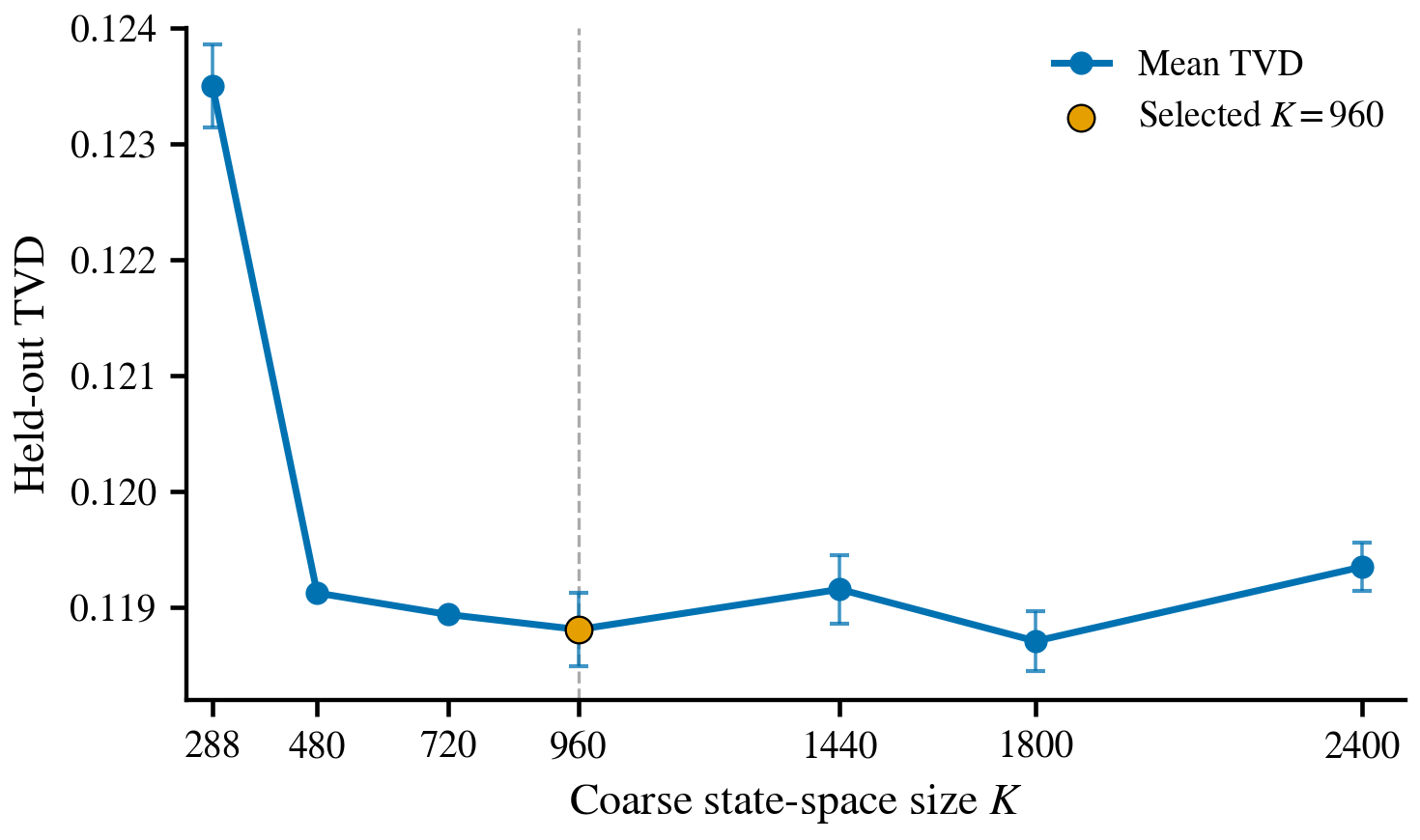}
\caption{Sensitivity of held-out TVD to the coarse state-space size $K$ under the POI+LODES condition setting. Points show mean TVD across seeds, and error bars show seed-level standard deviation when three seeds are available.}
\label{fig:s_k_sweep}
\end{figure}

\begin{table}[H]
  \caption{Held-out state checks for regional joint distribution recovery. Gain is computed as $\mathrm{TVD}_{\mathrm{IPF}}-\mathrm{TVD}_{\mathrm{hier}}$, so positive values indicate lower error for our proposed generative framework.}
  \label{tab:S3_state_holdout_checks}
  \centering
  \small
  \begin{tabular}{ccccc}
  \toprule
  \begin{tabular}[c]{@{}c@{}}Held-out \\ State\end{tabular} & PUMAs & \begin{tabular}[c]{@{}c@{}}Proposed Generative \\ Framework TVD\end{tabular} & \begin{tabular}[c]{@{}c@{}}IPF \\ TVD\end{tabular} & \begin{tabular}[c]{@{}c@{}}The Difference \\ of TVD\end{tabular} \\
  \midrule
  Michigan  & 68  & 0.1188 & 0.1270 & 0.0082 \\
  Florida   & 168 & 0.1378 & 0.1414 & 0.0036 \\
  Texas     & 217 & 0.1307 & 0.1403 & 0.0097 \\
  Wisconsin & 43  & 0.1277 & 0.1390 & 0.0114 \\
  \bottomrule
  \end{tabular}
\end{table}

\end{document}